\documentclass[12pt]{iopart} \usepackage{graphicx,subfigure}
\usepackage{cite,iopams} \renewcommand{\l}{\left}
\renewcommand{\r}{\right} \newcommand{\erf}{{\rm erf}}
\newcommand{\erfc}{{\rm erfc}}
\newcommand{\avg}[1]{\left\langle{#1}\right\rangle}
\newcommand{\ovl}[1]{\overline{#1}}

\newcommand{\bsy}[1]{\boldsymbol{#1}}
\newcommand{\beq}{\begin{equation}} \newcommand{\eeq}{\end{equation}}

\begin{document}

\title[Von Neumann's model for large random economies]
{Typical properties of optimal growth in the Von Neumann expanding
  model for large random economies}

\author{A De Martino\dag~ and M Marsili\ddag}

\address{\dag INFM-SMC and Dipartimento di Fisica, Universit\`a di
Roma ``La Sapienza'', p.le A Moro 2, 00185 Roma (Italy)}

\address{\ddag The Abdus Salam ICTP, Strada Costiera 11, 341014
Trieste (Italy)}

\ead{andrea.demartino@roma1.infn.it, marsili@ictp.it}

\begin{abstract}
  We calculate the optimal solutions of the fully heterogeneous Von
  Neumann expansion problem with $N$ processes and $P$ goods in the
  limit $N\to\infty$. This model provides an elementary description of
  the growth of a production economy in the long run. The system turns
  from a contracting to an expanding phase as $N$ increases beyond
  $P$. The solution is characterized by a universal behavior,
  independent of the parameters of the disorder statistics.
  Associating technological innovation to an increase of $N$, we find
  that while such an increase has a large positive impact on long term
  growth when $N\ll P$, its effect on technologically advanced
  economies ($N\gg P$) is very weak.
\end{abstract}

\section{Introduction}

The dynamics in many complex systems involves a flux through
components arranged in a heterogeneous network. Examples range from
cell metabolism \cite{methabolism}, food webs \cite{foodweb}, supply
networks \cite{supply}, and river networks \cite{rivers}, to the way
in which raw materials are combined and transformed into intermediate
or consumption goods in an economy \cite{JVN}.  Mathematically, these
phenomena can be cast into linear programming problems, for solving
which efficient algorithms exist even for large instances. In each of
these cases, however, it is important to understand whether the
emerging global properties are due to the specific wiring of the
input-output relationships or whether they are generic of large random
realizations of any such problem. Put differently, the study of the
typical properties of large random systems provides a useful ``null
hypothesis'' against which specific results should be compared. The
theoretical machinery for this kind of study is provided by the
statistical mechanics of disordered systems and has already been
applied to specific large random linear programming problems in the
past \cite{Knap,Nishi}.

One of the areas where such problems arise more frequently is
economics. The study of typical properties of large random economies
\cite{Follmer,Geneq} is particularly relevant, first because it allows
one to go beyond the simplistic framework of the so-called
`representative agent', by accounting for the heterogeneity across
agents and their interactions in many dimensions (technological
capabilities, budgets, endowments, etc).  Secondly because, contrary
to biology, economic time scales are such that evolutionary design
might not play a dominant role in shaping the (global properties of)
interaction network. Hence, in real economies the latter might be
closer to a large random instance than for e.g. metabolic networks.

In this paper, we will study a simple model of economic growth, put
forward by J. Von Neumann in the 1930's \cite{JVN}. It describes
economic growth as an autocatalytic process by which the outputs
generated at any time are used either for consumption or as inputs for
production at the later stage.  This model has been widely studied and
plays a central role in economic theory \cite{Champ,GaleKT,KMT,MT}, as
it forms the backbone of more refined models providing the key
insights for understanding growth in the long run\footnote{Models of
economic growth address the issue of maximizing discounted welfare
over the evolution paths. Turnpike theorems \cite{turnpike} show that
optimal paths overlap significantly with the paths of maximal
expansion described by Von Neumann's model.}. Our aim is that of
characterizing the growth properties in terms of the underlying
structural complexity of the production activity. More precisely, we
shall compute the maximal growth rate of the economy as a function of
the ratio between the number of production processes and the number of
goods, and of the parameters of the distribution of input-output
matrices. We shall also compute the number of active production
processes and of ``intermediate'' goods, whose output is used entirely
for further production. These results shed light on the way in which
growth is affected by technological innovation, namely by an
enrichment of the repertoire of available technologies, and in turn
how this affects the activity levels.

In the rest of the paper, we shall first introduce the model, then
present the statistical mechanics approach and finally discuss the
results. 

\section{The model}

In somewhat simplified terms, Von Neumann's expanding model may be
presented as follows. One considers an economy with $P$ commodities
(labeled $\mu$) and $N$ technologies (labeled $i$), each of which can
be operated at a non-negative scale $S_i\geq 0$ and is characterized
by an output vector $\boldsymbol{a}_i=\{a_i^\mu\}$ and by an input
vector $\boldsymbol{b}_i=\{b_i^\mu\}$, such that $S_i a_i^\mu$
(respectively $S_i b_i^\mu$) denotes the units of commodity $\mu$
produced (respectively used) by process $i$ when run at scale $S_i$.
It is assumed that input/output vectors are fixed in time and that
operation scales are the degrees of freedom to be set, for instance,
by firms. At time (or period) $t$, the economy is characterized by an
aggregate input and output vector for each commodity, $I^\mu(t)=\sum_i
S_i(t)b_i^\mu$ and $O^\mu(t)=\sum_i S_i(t)a_i^\mu$ respectively. Part
of the latter will be used as the input at period $t+1$ whereas the
rest, namely
\begin{equation}\label{cmu}
C^\mu(t)\equiv O^\mu(t)-I^\mu(t+1)
\end{equation}
is consumed at time $t$. In absence of external sources, a necessary
condition is that inputs at any time do not exceed the outputs at the
previous time, i.e. one must have $C^\mu(t)\geq 0$ for all $\mu$ at
all times. In this way, the model describes a closed economy which is
able to provide the society with all commodities without relying on
external sources. Modern economic growth theories introduce a value
for the stream of consumption $C^\mu(t)$ for all $t$ and $\mu$ --
usually by postulating a utility function and a discount factor -- and
look for optimal growth paths $\{S_i(t),~t\ge 0\}_{i=1}^N$, as a
function of some initial condition $I^\mu(0)$. Von Neumann's model
instead focuses on the simpler issue of studying the feasibility of
paths with a constant rate -- i.e. such that $I^\mu(t+1)=\rho
I^\mu(t)$ with $\rho>0$ a constant -- and, in particular, on computing
the highest feasible growth rate. These two problems are related because, 
under generic conditions, the optimal path coincides with that of
maximal expansion apart from an initial transient
\cite{turnpike}. This is why Von Neumann's model is relevant for
long run properties of models of economic growth. 
For paths with constant expansion rate, the
scales of production have the form $S_i(t)=s_i \rho^t$ and likewise
$C^\mu(t)=c^\mu \rho^t$. The (technological) expansion problem then
amounts to calculating the maximum $\rho>0$ such that a configuration
$\boldsymbol{s}=\{s_i\geq 0\}$ satisfying the condition
\begin{equation}\label{1}
c^\mu\equiv \sum_i s_i\l(a_i^\mu-\rho b_i^\mu\r)\geq 0~~~~~~~\forall\mu
\end{equation}
exists \cite{Gale,Thompson}.  In such a configuration the aggregate
output of each commodity is at least $\rho$ times its aggregate
input. If the maximum $\rho$, which we denote by $\rho^\star$, is
larger than 1 the economy is `expanding', whereas it is `contracting'
for $\rho^\star<1$. It is a rigorously established fact that if
$a_i^\mu\geq 0$ and $b_i^\mu\geq 0$ for all $i$ and $\mu$,
$\rho^\star$ exists (see \cite{Gale} for a simple proof). On the other
hand, the actual value of $\rho^\star$ is expected to depend on the
input and output matrices. Intuitively, $\rho^\star$ should increase
with the number $N$ of technologies and decrease when the economy is
required to produce a larger number $P$ of goods.

\section{Statistical mechanics of large random instances}

In this work we study a fully heterogeneous version of above problem
with technologies different from each other. We focus
on random instances where $a_i^\mu$ and $b_i^\mu$ are quenched random
variables drawn from a certain probability distribution. More
specifically, we shall consider the pair $(a_i^\mu,b_i^\mu)$ as
independent and identically distributed for each $i$ and $\mu$. To
begin with, let us first simplify the problem by writing
$a_i^\mu=\ovl{a}(1+\alpha_i^\mu)$ and
$b_i^\mu=\ovl{b}(1+\beta_i^\mu)$, where $\ovl{a}$ and $\ovl{b}$ are
positive constants while $\alpha_i^\mu,~\beta_i^\mu$ are zero-average
quenched random variables. Inserting these into (\ref{1}) one easily
sees that to leading order (in $N$) $\rho^\star$ is given by the ratio
$\ovl{a}/\ovl{b}$ of the average output and average input
coefficients. In particular, the leading part of $\rho$ does not
depend on the structure of the input output matrices. The non trivial
aspects of the problem are related to the corrections to the leading
part. We therefore write the growth rate as
\begin{equation}
\rho=\frac{\ovl{a}}{\ovl{b}}\left(1+\frac{g}{\sqrt{N}}\right)\label{rho}
\end{equation}
so that 
(\ref{1}) becomes
\begin{equation}
{c^\mu}=\bar{a}\sum_i s_i\left[\alpha_i^\mu-\frac{g}{\sqrt{N}}-
\left(1+\frac{g}{\sqrt{N}}\right)\beta_i^\mu\right]\geq 0~~~~~~~\forall\mu
\label{cimu}
\end{equation}
The problem thus reduces to that of finding the largest value
$g^\star$ of $g$ for which it is possible to find coefficients
$\{s_i\geq 0\}$ satisfying (\ref{cimu}). 

This issue can be tackled in the limit $N\to\infty$ employing a
standard technique originally due to Gardner \cite{Gardner}, which
allows to derive the behavior of $g^\star$ as a function of the
control parameter $n=\lim_{N\to\infty}N/P$.  The volume of
configuration space occupied by micro-states satisfying (\ref{1}) at
fixed disorder is given by
\begin{eqnarray}
\fl V_{\boldsymbol{\alpha,\beta}}(g)= \int_0^\infty d\boldsymbol{s}\prod_\mu
\theta \left[ \frac{1}{\sqrt{N}}\sum_i s_i\left[
\alpha_i^\mu-\frac{g}{\sqrt{N}}- \left(1+\frac{g}{\sqrt{N}}\right)
\beta_i^\mu \right]\r] \delta\left(\sum_is_i -N\right)
\end{eqnarray}
where we introduced a linear constraint $\sum_i s_i=N$. The typical
volume occupied by solutions for $N\to\infty$ reads instead
\[
V_{{\rm typ}}(g)\sim e^{Nv_{{\rm typ}}(g)}
\]
where
\begin{equation}\label{Vtyp}
v_{{\rm typ}}(g)=\lim_{N\to\infty}\frac{1}{N} \ovl{\log
V_{\boldsymbol{\alpha,\beta}}(g)}= \lim_{r\to
0}\lim_{N\to\infty}\frac{1}{Nr}\log\ovl{\l[
V_{\boldsymbol{\alpha,\beta}}(g)\r]^r}
\end{equation}
where the last equality contains the replica trick and the over-bar
stands for an average over the quenched disorder, that is over the
vectors $\alpha_i^\mu$ and $\beta_i^\mu$. As usual, the leading
contributions to $v_{{\rm typ}}(g)$ in the limit $N\to\infty$ come
from the first two moments of the distribution of $\alpha_i^\mu$ and
$\beta_i^\mu$. Given that $\alpha_i^\mu$ and $\beta_i^\mu$ have zero
mean, the only property of the disorder distribution which enters the
final result is the covariance matrix of the disorder. Actually, the
explicit calculation shows that $g^\star$ only depends on the
parameter
\begin{equation}\label{k}
 k=\ovl{(\alpha_i^\mu-\beta_i^\mu)^2}
\end{equation}
To make a concrete example, consider the input (output) matrices where
$b_i^\mu=b$ ($a_i^\mu=a$) for $B$ values of $\mu$ and $b_i^\mu=0$
($a_i^\mu=0$) otherwise. Then it is easy to see that $\ovl{b}=bB/N$
($\ovl{a}=aA/N$), $\ovl{(\beta_i^\mu)^2}=N/B-1$
($\ovl{(\alpha_i^\mu)^2}=N/A-1$) and $\ovl{\alpha_i^\mu\beta_i^\mu}=0$
so that $k=N/A+N/B-2$. In particular, the case where outputs and
inputs are few corresponds to $k\gg 1$.

After expressing the $\theta$-functions via their integral
representations and carrying out the disorder average one finds
\begin{equation}
\ovl{[V_{\boldsymbol{\alpha,\beta}}(g)]^r}=\int J_1(\bsy{q})
J_2(\bsy{q})d\bsy{q}
\end{equation}
where $\bsy{q}$ is a vector of order parameters
$\{q_{\ell\ell'}\}_{\ell\leq\ell'}^{1,r}$ representing the overlaps
between the configurations in different replicas,
$q_{\ell\ell'}=(1/N)\sum_i s_{i\ell}s_{i\ell'}$, and
\begin{eqnarray}
J_1(\bsy{q})&=&\int_0^\infty D\bsy{c} \int_{-\infty}^\infty D\bsy{z} \prod_\mu  e^{i\sum_\ell
   z^\mu_\ell (c_\ell^\mu+g)-\frac{k}{2}\sum_{\ell,\ell'} q_{\ell\ell'}z^\mu_\ell
   z^\mu_{\ell'}} \label{uns}\\
   J_2(\bsy{q})&=&\int_0^\infty D\bsy{s}\prod_\ell \delta\left(\sum_i s_{i\ell} -N\right)
  \prod_{\ell\leq\ell'}\delta\left(\sum_i s_{i\ell}s_{i\ell'}
   -Nq_{\ell\ell'}\right)\nonumber\\
&=&\int d\bsy{R} d\bsy{m}\int_0^\infty D\bsy{s} ~
e^{-\sum_\ell m_\ell(\sum_i s_{i\ell}-N)-\sum_{\ell\leq \ell'}R_{\ell\ell'}(\sum_i s_{i\ell}s_{i\ell'}-Nq_{\ell\ell'})}\label{dus}
\end{eqnarray}
The space of solutions $\{s_i\}$ is a convex set, hence we expect the
replica-symmetric approximation to be exact in this case. We therefore
evaluate (\ref{uns}) and (\ref{dus}) imposing the Ansatz
\begin{equation}\label{rsapprox}
q_{\ell\ell'}=q+\chi \delta_{\ell\ell'},~~~~~
R_{\ell\ell'}=\frac{\beta+\tau^2}{2}\delta_{\ell\ell'}-\tau^2, ~~~~m_\ell=m.
\end{equation}
Putting things together, Eq. (\ref{Vtyp}) yields
\begin{equation}
v_{{\rm typ}}(g)={\rm extr}_{q,\chi}\left[F_1(q,\chi)+
{\rm extr}_{\beta,\tau,m}F_2(q,\chi,\beta,\tau,m)\right]
\end{equation}
where ${\rm extr}_{x} f(x)$ denotes the operation of taking the
extremum of $f(x)$, and 
\begin{eqnarray}
F_1&=\frac{1}{n}\avg{\log\int_0^\infty\frac{dc}{\sqrt{2\pi k\chi}}
e^{-\frac{(c+g+\xi\sqrt{kq})^2}{2k\chi}}}_\xi\label{bos}\\
&=\frac{1}{n}\avg{\log\l[\frac{1}{2}\erfc\frac{g+\xi\sqrt{qk}}{\sqrt{2k\chi}}\r]}_\xi\nonumber\\
F_2&=m+\frac{1}{2}\beta(\chi+q)-\frac{1}{2}\chi\tau^2+\avg{\log\int_0^\infty\label{boss}
ds~e^{-(m+\xi\tau)s-\beta s^2/2}}_\xi\label{f2}
\end{eqnarray}
The brackets
$\avg{\ldots}_\xi=\int_{-\infty}^\infty\frac{d\xi}{\sqrt{2\pi}}
e^{-\xi^2/2}\ldots$ stand for an average over the unit
variance Gaussian random variable $\xi$. The problem can in principle be solved
in a straightforward way, by analyzing the saddle point equations for
any value of $g$. We expect, however that as $g\to g^\star$ the
typical volume $V_{{\rm typ}}$ shrinks until just one solution remains
(modulo re-scalings of the $s_i$'s). Hence $\chi$, which describes the
fluctuation $s_i$ among feasible solutions, should also vanish as
$g\to g^\star$. Hence the conditions $g=g^\star$ and $\chi=0$ are
equivalent. When $\chi\to 0$ the integral in (\ref{bos}) can be
calculated by steepest descent. The distribution of $c^\mu$ in Eq.
(\ref{cimu}) can be read off Eq. (\ref{bos}) and it has the form
\begin{equation}
p(c)=\phi_0\delta(c)+\theta(c)e^{-\frac{(c+g^\star)^2}{2q k}}
\end{equation}
where 
\begin{equation}
\phi_0=\frac{1}{2}\l(1+\erf\frac{g^\star}{\sqrt{2qk}}\r)\label{phi0}
\end{equation}
is the fraction of commodities that are not consumed (namely for which
$c^\mu=0$) at $g=g^\star$. These commodities are uniquely used as
intermediate goods in the production process. A further important
observation is that $F_1$ and hence $v_{{\rm typ}}$ depend on
$g^\star$ only through the combination $g^\star/\sqrt{k}$. This
completely characterizes the dependence of the maximal growth rate on
the parameters of the disorder distribution.

If as $g\to g^\star$ the volume of feasible solutions shrinks to zero,
the integral on $s$ in $F_2$ must ultimately be dominated by a single
value. This is consistent with a $1/\chi$ divergence of the parameters
$\beta,~\tau$ and $m$. Hence it is convenient to introduce the
variables
\begin{equation}
b=\chi\beta,~~~~~t=\chi\tau,~~~~~z=-m/\tau
\end{equation}
which remain finite in the limit $\chi\to 0$ and turn out (after some
algebra) to be given by the solution of
\begin{equation}
\eqalign{
q=\frac{\avg{(z-\xi)^2\theta(z-\xi)}_\xi}{\avg{(z-\xi)\theta(z-\xi)}_\xi^2}\\
b=-\avg{\xi(z-\xi)\theta(z-\xi)}_\xi\\
t=-\frac{\avg{\xi(z-\xi)\theta(z-\xi)}_\xi}{\avg{(z-\xi)\theta(z-\xi)}_\xi}}.\label{yas}
\end{equation}
In analogy with what we did for $c^\mu$, it is possible to compute the
distribution of $s_i$. In particular the fraction of inactive
technologies ($s_i=0$) is found to be 
\beq
\psi_0=\frac{1}{2}\erfc\left(\frac{z}{\sqrt{2}}\right).  
\eeq
Eqs. (\ref{yas}) can be solved numerically to the desired accuracy to
yield $b$, $t$ and $z$ as a function of $q$.  We are left with two
saddle point conditions, for $q$ and $\chi$, respectively. The latter
reads 
\beq\label{chisp}
t^2=\frac{q}{n}\avg{\l(\xi+g/\sqrt{kq}\r)^2\theta\l(\xi+g/\sqrt{kq}\r)}_\xi
\eeq 
The saddle point equation for $q$ requires a bit more
work. Indeed for $\chi\to 0$ the leading contributions of $F_1$ and
$F_2$ are of order $1/\chi$ but they cancel exactly. One must
therefore consider the next-to-leading-order correction, of order
$\log \chi$. We note, in passing, that this is consistent with the
volume of solutions vanishing as $v_{{\rm typ}}(g)\sim
\chi^{\gamma N}$ as $g\to g^\star$. After some algebra, the final equation
takes the particularly simple form \beq
\phi_0=n(1-\psi_0).\label{psiz} \eeq This condition implies that the
number of active processes equals that of intermediate commodities at
$g^\star$. Noting that for any $\mu$ such that $c^\mu=0$ we have a
linear equation for the scales $s_i>0$, we see also that (\ref{psiz})
simply corresponds to the requirement that the number of equations
should match the number of variables.

Eqs. (\ref{yas}), (\ref{chisp}) and (\ref{psiz}) allow us to compute
$g^\star$ as a function of $n$ and $k$. As already noticed, the
dependence $g^\star\propto \sqrt{k}$ can be read off directly from the
equations, hence $g^\star/\sqrt{k}$ is an universal function of $n$,
independent of the details of the distribution of input-output
coefficients (as long as they are i.i.d.). Actually, since $k$ enters
the equations only through the combination $g^\star/\sqrt{k}$, and this is
a function of $n$, then the solution itself will be universal,
i.e. independent of $k$.

\begin{figure}
\begin{center}
\includegraphics*[height=7cm]{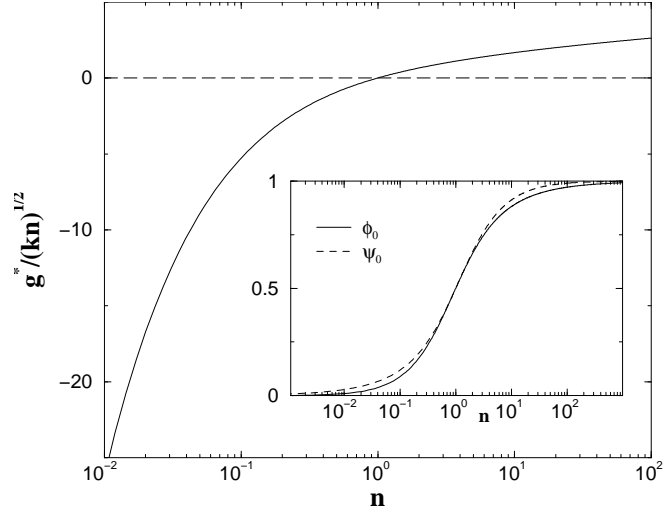}
\end{center}
\caption{\label{uno}Behavior of $g^\star/\sqrt{kn}$ vs $n$. 
Inset: $\phi_0$ and $\psi_0$ (related by (\ref{psiz})) vs $n$.} 
\end{figure}

Fig. \ref{uno} reports the rescaled growth rate $g^\star/\sqrt{kn}$ as a
function of $n$. This line separates the region of feasible solutions
with growth rates $g\le g^\star$ from the region of unfeasible solutions.
$g^\star$ crosses the line $g^\star=0$ at $n=1$, as can be checked explicitly.
Indeed, by Eq. (\ref{phi0}), $g^\star=0$ corresponds to a situation where
half of the goods are not consumed ($\phi_0=1/2$) while Eq.
(\ref{chisp}) yields $t^2=q/(2n)$, which is consistent with the
other equations for $z=0$, i.e. $n=1$. At this same point, half of the
technologies are operated ($\psi_0=1/2$).  This means that, when there
are more technologies than goods ($n>1$) a growth rate higher than
that of the ratio $\ovl{a}/\ovl{b}$ of average output and input
coefficients is achievable. The growth rate is instead smaller than
$\ovl{a}/\ovl{b}$ when $n<1$.

The inset of Fig. \ref{uno} shows the fraction of inactive processes
$\psi_0$ and that of intermediate commodities $\phi_0$ at $g^\star$, as a
function of $n$. For what we said earlier, these are universal
functions of $n$ independent of the details of the disorder
distribution.  Both $\phi_0$ and $\psi_0$ tend to one when $n$
increases. 

\section{Discussion}

In summary, we have studied the typical properties of the Von Neumann
expanding model in the case where input-output matrices have i.i.d.
random elements. We characterize the region of feasible expansion
paths and focus on the solutions at its boundary, which correspond to
the paths of maximal expansion.  We uncover an universal behavior,
independent of the details of the distribution, of the relevant
quantities. In particular we find that as the number $N$ of
technologies grows, the optimal growth rate increases, but the economy
becomes more and more selective both on the processes which are used
and on the goods which are consumed.

At a purely speculative level, our results allow us to draw
conclusions on how long term growth on the maximal expansion path will
be affected by technological innovation. The latter, defined as the
introduction of new designs, i.e. new feasible ways of combining
inputs to produce desirable outputs \cite{Romer} would just
correspond, in our simplified world, to an increase in the number $N$
of transformation processes which the economy has at its disposal. By
Eq. (\ref{rho}) the change in the growth rate is related to the change
in $g^\star/\sqrt{n}$, which is plotted in Fig. \ref{uno}. This shows
that when $n$ is small ($n\ll 1$)
\[
\delta\rho\propto\frac{\delta n}{n^{3/2}\sqrt{P}}
\]
i.e. an increase in $N$ can have a large positive impact on long term
growth. For technologically mature economies ($n\gg 1$) instead,
$g^\star/\sqrt{n}$ increases much more slowly, hence technological
innovation has much smaller effect on long term growth. These insights
are remarkably similar to those derived in \cite{Geneq} for model of
general economic equilibrium.

There are several directions in which the present work could be
extended. First, it would be desirable to study more realistic
ensembles of input-output matrices, or more detailed models of
economic growth. In particular, realistic technologies only have a
finite number of inputs and outputs, which would call for the use of
techniques for disordered systems with diluted interactions. It would
also be interesting to generalize this study to the case where the
input-output transformation is subject to stochasticity. A further
natural extension of this approach concerns the analysis of typical
properties of large random metabolic \cite{methabolism} or supply
networks \cite{supply}, or of food webs \cite{foodweb}.

\end{document}